\documentclass[twocolumn,aps,floatfix,superscriptaddress,showpacs,letterpaper]{revtex4}
\usepackage{amsmath,amssymb,eucal,graphicx}

\begin{document}
\title{Logarithmic current fluctuations in non-equilibrium quantum spin chains}

\author{T. Antal}
\affiliation{Program for Evolutionary Dynamics, Harvard University, Cambridge, MA 02138, USA}
\author{P.~L.~Krapivsky}
\affiliation{Department of Physics, Boston University, Boston, MA 02215, USA}\affiliation{Theoretical Division and Center for
Nonlinear Studies, Los Alamos National Laboratory, Los Alamos, New
Mexico 87545, USA}
\author{A.~R\'akos}
\affiliation{Research Group for Condensed Matter Physics of the Hungarian Academy of Sciences, H-1111 Budapest, Hungary}

\begin{abstract}
We study zero-temperature quantum spin chains which are characterized by a non-vanishing current. For the $XX$ model starting from the initial state $|\cdots \uparrow\uparrow\uparrow\downarrow\downarrow\downarrow\cdots\rangle$ we derive an exact expression for the variance of the total spin current. We show that asymptotically the variance exhibits an anomalously slow logarithmic growth; we also extract the sub-leading constant term. We then argue that the logarithmic growth remains valid for the $XXZ$ model in the critical region.   
\end{abstract}

\pacs{05.60.Gg, 75.10.Jm, 05.70.Ln}

\maketitle

\section{Introduction}

A distinguishing feature of non-equilibrium states is the presence of currents \cite{{ligget85},{schmittmann95},{schutz00}}. Fluctuations of currents often exhibit universal behavior \cite{Prahofer2000} and shed light on the nature of non-equilibrium 
systems.  Current fluctuations in classical systems have been extensively investigated, see e.g. \cite{Ferrari1994, Derrida1998, Lebowitz1999, Johansson2000, Prahofer2000, bodineau04, harris05, Sasamoto2005a, Ferrari2006, Rakos2006a, PK} and a review \cite{Derrida2007}.  The total current grows linearly with time and current fluctuations usually exhibit an algebraic growth. Quantum fluctuations in general, and spin current fluctuations in particular, are much less understood; even in the simplest systems quantum fluctuations often behave very differently from standard statistical fluctuations (see e.g. \cite{YC}). 

In this paper we study current fluctuations in quantum spin chains.  How to impose currents in spin chains? Perhaps the simplest way is to start with a spin chain in the following inhomogeneous product state
\begin{equation}
\label{initial}
|\cdots \uparrow\uparrow\uparrow\downarrow\downarrow\downarrow\cdots\rangle
\end{equation}
This choice  \cite{spin99} allows one to avoid complications and arbitrariness of coupling the chain to spin reservoirs. 

State \eqref{initial} evolves according to the Heisenberg equations of motion. The average magnetization profile has been computed for the simplest quantum chains, e.g. for the $XX$ model where the perturbed region was found to grow ballistically \cite{spin99}. Numerical works \cite{gobert05} suggest that the growth is also ballistic for the $XXZ$ chain in the critical region (described by Hamiltonian \eqref{hami} with $|\Delta|<1$).

The main goal of this paper is to study the fluctuations of spin current in quantum chains. After a brief description of the model in section \ref{model}, in section \ref{hidro} we present a simple derivation of the asymptotic properties of non-equilibrium states in free fermion systems with special initial conditions. In section \ref{fluct} we probe fluctuations of the current specifically in the $XX$ spin chain. First we present a back-of-the-envelope calculation for the variance of the time integrated current; then we establish an exact result from which we extract the long time asymptotical behavior. We discuss the more general $XXZ$ chain in section \ref{XXZ}. A summary of our results and relation to other work is given in section \ref{discussion}.

\section{Model}
\label{model}

Most generally, we consider the quantum $XXZ$ Heisenberg spin chain with Hamiltonian 
\begin{equation}
\label{hami}
\mathcal{H}=-\sum_{n} \left(s^x_n s^x_{n+1}
+ s^y_n s^y_{n+1} + \Delta s^z_n s^z_{n+1}\right) 
\end{equation}
Here we set the coupling constants to unity in the $x$ and $y$ directions. The coupling constant $\Delta$ in the $z$ direction is called the anisotropy parameter. The $z$ component of the total magnetization $M^z=\sum_{n=-\infty}^\infty s^z_n$ is a conserved quantity in this model, and our aim is to study the corresponding current. In the following we shall focus on the time evolution of this spin chain starting from an inhomogeneous initial state, whereby the left and right halves of the infinite chain are set to different quantum states and are joined at time zero. Such initial states provide a particularly convenient framework to study currents and their fluctuations in quantum spin chains. 

We mainly consider the special case of $\Delta=0$, where the model reduces to free fermions. In this system, known as the $XX$ model, the time evolution can be written in a compact form, which enables us to perform exact calculations. In particular, it is possible to evaluate the scaling limit of the magnetization profile and other physical quantities \cite{spin99,ogata02}. Corrections to this scaling behavior were considered in \cite{{karevski02},{hunyadi04}}. 

Interesting non-equilibrium behavior was found in disordered spin chains \cite{abreit02,platini05} and chains at finite temperatures \cite{ogata02,ogata02b,platini07}.
An alternative method has been proposed to generate stationary currents in spin chains using a Lagrange multiplier \cite{spin97,spin98,cardy00,kosov04,racz00}. Using this method fluctuations of basic quantities have been studied in non-equilibrium steady states in \cite{eisler03}.

The relaxation from a large class of initial conditions was considered in numerous studies starting from late sixties; see \cite{niemeijer67,tjon70,berouch69,igloi00} and \cite{berim02} for a review of more recent work. 
Our focus, however, is on non-equilibrium states with a non-vanishing current.

\section{Hydrodynamic description}
\label{hidro}

We begin by describing a simple method that allows one to obtain the long time asymptotic behavior of a free fermion system by employing a continuous hydrodynamic description. This helps to avoid a lengthy exact calculation and yet the final results are asymptotically exact. Specifically, a justification of this approach for the $XX$ model is given by exact results \cite{spin99,ogata02}, which yield the same asymptotical behavior as the hydrodynamical description discussed below. 

The Hamiltonian of the free fermion system can be written in the form
\begin{equation}
\mathcal{H}=\sum_k \epsilon(k) \eta_k^\dag \eta_k 
\end{equation}
where $\eta^\dag_k$ and $\eta_k$ are creation and annihilation operators of fermions with momentum $k$ and $\epsilon(k)$ is the energy of an excitation with wave number $k$.
 
In the simplest situation, the system is initially divided into two half infinite chains, each of them being in a homogeneous pure state. In this case, the elementary excitations can be considered initially homogeneously distributed in each half chains. At time zero, each mode starts moving with velocity $v(k)=\epsilon'(k)$. As the excitations are entirely independent, they do not interact and keep moving with their initial velocities. This argument suggests that whether an excitations is present at a space-time point $(n,t)$ depends only on the ratio $x=n/t$. Moreover, keeping $x$ fixed, a finite neighborhood of site $n$ becomes asymptotically homogeneous for $t\to\infty$. This physical picture is not exact due to the finite lattice spacing. However, we believe that the above description becomes asymptotically exact for any free fermion system in the scaling limit: $n\to\infty, t\to\infty, ~n/t=\mathrm{const}$. (See \cite{ogata02b} for a rigorous derivation of this scaling limit for the $XX$ model.)

Whether an excitation is present at position $n$ at time $t$ can be decided by noting that for $n>0$,  the modes which are present  were initially on the left side of the chain with $v(k)>n/t$ and on the right side of the chain with $v(k)<n/t$. Similar argument applies for $n<0$. 

This method can be extended to the general case when the two half-infinite chains are initially in mixed states, e.g., one can consider the situation when the two half-infinite chains are set to different temperatures \cite{ogata02b}.

\begin{figure}
\includegraphics{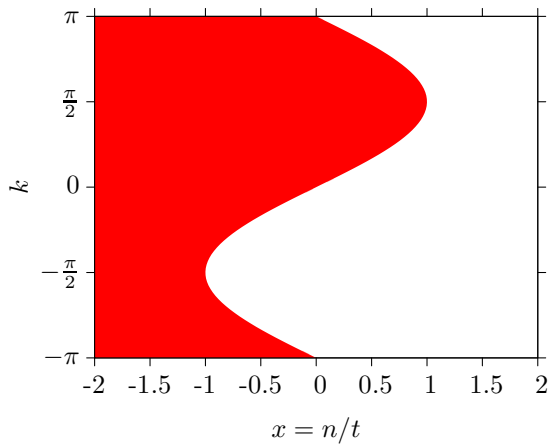}
\caption{(Color online) Hydrodynamic description for the $XX$ chain started from the $|\cdots \uparrow\uparrow\uparrow\downarrow\downarrow\downarrow\cdots\rangle$ initial condition. We consider the scaling limit where $t\to\infty$ and $x=n/t=\text{const}$. The shaded region shows the elementary excitations that are present in the scaling points indexed by $x$.}
\label{fig:XX_hydro}
\end{figure}

As an illustration, let us calculate the magnetization profile in the $XX$ model with the simplest  initial condition \eqref{initial}. The spectrum of the model is $\epsilon(k)=-\cos(k)$, that is, $v(k)=\sin(k)$. Initially, all the modes are filled on the left, while the right side of the chain is in the vacuum state. At time $t$, around site $n>0$, the modes with $\sin(k)>n/t$ are filled, that is, all the modes with $k_0<k<\pi-k_0$, where $k_0=\arcsin(n/t)$. Similarly, for $n<0$, only the modes with $\sin(k)>n/t$ are filled, that is,  the modes with $-\pi<k<-\pi-k_0$, and $k_0<k<\pi$. For an illustration see Fig.~\ref{fig:XX_hydro}. As each mode carries a unit magnetization, the average $z$ magnetization can be obtained   by simply integrating through the filled modes $m(x=n/t) = 1/2 + \int dk/2\pi$. This results in the well known profile $m(x=n/t) = -\frac{1}{\pi} \arcsin(x)$ for $-1<x<1$, and the magnetization keeps its initial values outside this region. This limiting profile was obtained in \cite{spin99} by exact calculation.

Other applications are given in Appendix \ref{examples}. 

\section{Fluctuations of the Current}
\label{fluct}

The local magnetization current operator for a quantum spin chain can be obtained through a continuity equation for the local magnetization \cite{spin98}. For the $XX$ model this gives 
\begin{equation}
 j_n = s^y_n s^x_{n+1} - s^x_n s^y_{n+1}
\end{equation} 
for the current between spin $n$ and $n+1$. (We measure time in units of $\hbar$).
The time integrated current $C_0$, i.e., the net transported magnetization up to time $t$ through the bond between spin 0 and spin 1, is a quantity which is less obvious to define for a quantum system in general. However, in the case of the setup (\ref{initial}), the integrated current $C_0$ can be expressed in a simple way 
\begin{equation}
 C_0=\sum_{n\geq 1} (s^z_n+1/2).
\end{equation} 
The average of the integrated current $\langle C_0 \rangle$ through the central bond grows asymptotically as $\pi^{-1}t$ \cite{spin99}; alternatively, this can be seen from the hydrodynamic picture described above. 

The hydrodynamic approach of Sec.~\ref{hidro} does not allow to probe fluctuations. Therefore we must return to the microscopic description. Below we shall focus on the variance of the total current $D(t)\equiv \langle C_0^2\rangle -\langle C_0\rangle^2$.


Let us define the left, right and total magnetization as follows:
\begin{gather}\label{def}
M_L = \sum_{n\le 0} s_n^z, \qquad M_R = \sum_{n\ge 1} s_n^z, \cr M = M_L+M_R.
\end{gather}
The variance of the integrated current is equal to the variance of the left (right) magnetization:
\begin{equation}\label{egy}
D(t) = \langle M_R^2\rangle_t -\langle M_R \rangle_t^2 = \langle M_L^2\rangle_t -\langle M_L \rangle_t^2.
\end{equation} 
Since the total magnetization is conserved and the initial state is an $M$ eigenstate, the fluctuation of $M$ remains zero for any time $t$
\begin{equation}\label{harom}
\langle(M_L+M_R)^2\rangle_t -\langle M_L+M_R\rangle_t^2=0.
\end{equation}
By exploiting this property we can rewrite (\ref{egy}) as
\begin{multline}
\label{negy}
D(t)= \langle M_L \rangle_t \langle M_R \rangle_t - \langle M_L M_R \rangle_t \\
= \sum_{l\leq 0, m\ge1} \left( \langle s_l^z \rangle_t \langle s_m^z \rangle_t - \langle s_l^z s_m^z \rangle_t \right).
\end{multline}

Before presenting an exact calculation we provide a back-of-the-envelope derivation of our main result. The idea is to evaluate $D(t)$ by substituting correlations in \eqref{negy} with their stationary values in the local state which builds up at the origin for $t\to\infty$ (see \cite{spin99} and our Fig.~\ref{fig:XX_hydro}). The reason is that the main contribution comes from those spins for which $(l-m)$ is not too large, and for $t\gg 1$ these ``points'' are located near the origin. This leads to
\begin{equation}\label{ot}
D(t) = - \sum_{n>0} n \rho^z(n),
\end{equation}
where 
\begin{equation}\label{hat}
\rho^z(n) = \langle s_k^z s_{k+n}^z \rangle - \langle s_k^z \rangle \langle s_{k+n}^z \rangle.
\end{equation}
In the (homogeneous) ``maximal current'' stationary state of the $XX$ model, the correlator $\rho^z(n)$ takes the same form as in the ground state \cite{spin98} where it is given by the well-known expression:
\begin{equation}\label{het}
\rho^z(n) = 
\begin{cases}
-\frac{1}{\pi^2 n^2} & n=\text{ odd} \\
0 & n=\text{ even} \end{cases}.
\end{equation}
This gives a logarithmical divergence for $D(t)$, which one can regularize by truncating the sum in (\ref{ot}). For a finite but large $t$ the volume of the region around the origin, which can be described by this maximal current state, grows linearly with $t$. Hence we choose the upper limit in the sum in (\ref{ot}) to be proportional to $t$ and obtain
\begin{equation}\label{nyolc}
D(t) \sim - \sum_{n=1}^{\sim t} n \rho^z(n) = \frac{1}{2\pi^2}\ln(t).
\end{equation}  
In this expression, the factor $1/2$ appears since correlations between evenly spaced sites vanish.

This argument remains valid for a more general class of initial conditions, where on the left (right) half of the chain the fermions are filled up to the Fermi energy $\mu_L$ ($\mu_R$). In this case, the asymptotic state --- which builds up near the origin --- includes fermions with momenta  varying from $-k_R$ to $k_L$, where $k_L$ and $k_R$ are the Fermi momenta corresponding to $\mu_L$ and $\mu_R$. (For an illustration see Fig.~\ref{fig:XX_hydro2}; more details are given in Appendix \ref{examples_XX}.) The correlation function $\rho^z(n)$ for these asymptotic states can easily be calculated \cite{spin98}. One finds that in general it behaves as $\rho^z(n) = -\frac{1}{\pi^2 n^2} \sin^2(n\varphi)$, where $\varphi=(k_L+k_R)/2$ \cite{spin98}. As the $\varphi$ dependence of the asymptotic form averages out we conclude that the result \eqref{nyolc} is unchanged for this class of initial states.

The exact evaluation of $D(t)$ is based on (\ref{negy}). Following the strategy of \cite{spin99} we write $s_i^z$ in terms of the local fermionic creation and annihilation operators $c^\dagger, c$ as
\begin{equation}\label{siz}
s_n^z=c_n^\dagger c^{}_n -\frac{1}{2}.
\end{equation} 
In the Heisenberg picture, the time dependence of these operators, under the dynamics of the $XX$ chain, has a simple form 
\begin{equation}
\label{ct}
c_n(t) = \sum_{j=-\infty}^\infty i^{j-n} J_{j-n}(t) c_j,
\end{equation}
where $J_{n}(t)$ are the Bessel functions. Inserting this into (\ref{negy}) one gets
\begin{multline}
D(t) = \sum_{l\leq 0, m\ge1} \sum_{\alpha, \beta, \gamma, \delta} i^{-\alpha+\beta-\gamma+\delta} \\ 
\times J_{\alpha-l}(t) J_{\beta-l}(t) J_{\gamma-m}(t) J_{\delta-m}(t) \\
\times \left( \langle c_\alpha^\dagger c_\beta \rangle \langle c_\gamma^\dagger c_\delta \rangle - \langle c_\alpha^\dagger c_\beta c_\gamma^\dagger c_\delta \rangle \right).
\end{multline}
The expectation value in the above formula is taken in the initial state. One finds 
\begin{multline}\label{expextation}
\langle c_\alpha^\dagger c^{}_\beta \rangle \langle c_\gamma^\dagger c_\delta \rangle - \langle c_\alpha^\dagger c_\beta c_\gamma^\dagger c_\delta \rangle = \\ 
\begin{cases} - \delta_{\alpha, \delta} \delta_{\beta, \gamma}   & \text{if } \alpha\uparrow, \beta\downarrow \\ 0 & \text{otherwise} \end{cases}
\end{multline}
for initial states which are product states of individual spins pointing either up or down (like \eqref{initial}). Here, $\alpha\uparrow$, $\beta\downarrow$ are shorthand notations for $\alpha,\beta$ with $s_\alpha=\uparrow, s_\beta=\downarrow$.
Using identities $J_{-k}(t)=(-1)^k J_k(t)$ and \cite{gradshteyn}
\begin{equation*}
\sum_{k\ge1} J_{k+p}(t) J_{k+q}(t) =
t\, \frac{J_p(t)J_{q+1}(t) - J_{p+1}(t)J_q(t)}{2(p-q)}
\end{equation*}
for the sums over $l$ and $m$, one obtains 
\begin{equation}
\label{DJJ}
D(t)=\frac{t^2}{4} \sum_{\alpha\uparrow, \beta\downarrow} 
\left[
\frac{J_{\alpha-1}(t) J_{\beta}(t) - J_\alpha(t) J_{\beta-1}(t)}{\alpha-\beta}\right]^2.
\end{equation}
This expression for our initial condition \eqref{initial} becomes
\begin{equation}
\label{Dt-exact}
D(t)=\frac{t^2}{4}\sum_{l,m\geq 1}\left[\frac{ 
J_{l-1}(t) J_{m-1}(t) + J_l(t) J_m(t)}{l+m-1}\right]^2.
\end{equation}
This is an {\em exact} expression for the variance of the current which is valid at any time $t\geq 0$. From this formula we have deduced the long time asymptotic behavior 
\begin{equation}
\label{Dt-asymp}
D(t)=\frac{1}{2\pi^2}(\ln t + C)
\end{equation}
with constant $C = 2.963510026\ldots$. The leading term coincides with our heuristic argument \eqref{nyolc}. The derivation of  \eqref{Dt-asymp} is relegated to appendix \ref{long}.

\begin{figure}[htb]
\includegraphics{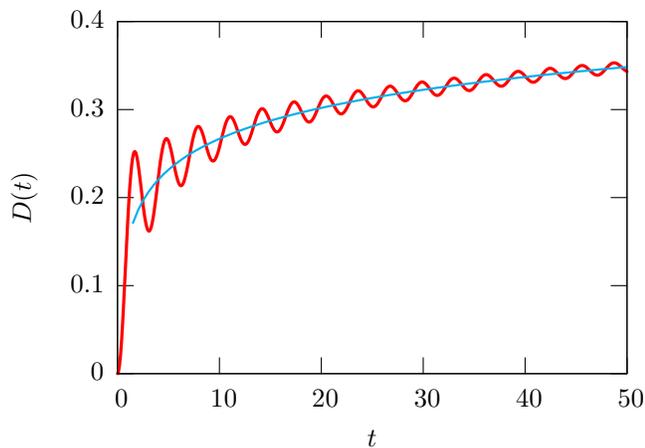}
\caption{(Color online) Shown is the variance $D(t)$ vs.\ time $t$. Red curve is a result of the exact numerical evaluation of \eqref{Dt-exact}, blue curve shows the result \eqref{Dt-asymp} of our asymptotical analysis. On top of the logarithmic growth we find a subleading oscillating term. Eq.~\eqref{improved_fit} gives a fit, which is almost indistinguishable from the numerical data (red line).}
\label{dtime}
\end{figure}

We also evaluated expression \eqref{Dt-exact} numerically and plotted it in Fig.~\ref{dtime} for $t<50$. We observe a logarithmic increase in time. The numerical estimate for the constant, $C\approx 2.9633$, is in a good agreement with the exact result. 

In addition to this logarithmic growth, one observes oscillations with decreasing amplitude. We found that the formula 
\begin{equation}
\label{improved_fit}
 D(t)=\frac{1}{2\pi^2}\left[ \ln t + C - \frac{\cos 2t }{t}\left( \ln t + C' \right) \right],
\end{equation} 
with $C'\approx 1.95$ (and $C$ given exactly in \eqref{C}) gives a very good fit to the numerical data even for relatively short times. Since the mean current is $1/\pi$, in average one fermion crosses the origin in time $\pi$. Hence the $\cos 2t$ oscillations can be interpreted as a consequence of the quantum nature of the magnetization: each passing fermion causes a bump in the fluctuations. Similar arguments were used to explain oscillations in the magnetization profile in the same system \cite{hunyadi04}.

\section{$XXZ$ spin chain}
\label{XXZ}

In a general $XXZ$ chain \eqref{hami}, the coupling is nonzero in the $z$ direction. There is no explicit solution for the time evolution in this general case. Due to the symmetry in the $XY$ plane, however, the $z$ component $M_z$ of the total magnetization is still conserved, hence the magnetization current can be studied. It has been investigated numerically by Gobert {\it et. al.} \cite{gobert05}. 

Our main interest lies in the so-called critical region ($-1<\Delta<1$); we shall also discuss the isotropic ferromagnetic ($\Delta=1$) and anti-ferromagnetic ($\Delta=-1$) spin chains. The $XX$ model belongs to the critical region, and the behavior in the entire critical region is believed to be similar to the behavior of the $XX$ model. In particular, the magnetization profile plausibly scales linearly with time: $m(n,t) \to \mathcal{M}(n/t)$ \cite{gobert05}.  On the other hand, due to the existence of kink-like ground states \cite{alcaraz95,matsui96}, the magnetization profile is expected to become frozen when $|\Delta|>1$ \cite{gobert05}. Algebraic scaling $m(n,t)\to \mathcal{M}(n/t^a)$ seems to emerge for the $XXX$ model ($|\Delta|=1$) in the scaling regime $n\to\infty$, $t\to\infty$ with $n/t^a$ kept finite. Numerically the exponent is $a=0.6\pm 0.1$ \cite{gobert05}, so the non-trivial part of the profile is
sub-ballistic. 

For the $XX$ spin chain, we have obtained the correct current fluctuations in the leading order from the simple formula \eqref{ot} when the upper limit of the sum was chosen to grow linearly with time. The reason for this choice of  upper limit is that the front and the whole profile ``moves'' linearly with time, hence the cutoff must behave similarly. We shall use \eqref{ot} also for the $XXZ$ chain. We shall assume that the upper bound moves linearly, namely as $vt$, in the critical region ($-1<\Delta<1$). The actual value of the `velocity' $v$ is unknown, but it does not affect the leading order term anyway. 

The next issue is whether one can use the equilibrium spin correlations $\rho^z(n)$ in the {\it presence} of current. For the $XX$ chain we know \cite{spin98,spin99} that current does not affect the $z$ component of the correlations significantly (may introduce a modulation), only the $x$ and $y$ components. Here we boldly assume the same for the $XXZ$ model, at least for its large distance behavior. Thus we use the equilibrium correlations in \eqref{ot}. The asymptotic formulae for $\rho^z(n)$ are \cite{lukyanov99,giamarchi04}
\begin{equation}
\label{rn}
\rho^z(n)\!=\!
\begin{cases}
-\delta n^{-2}                                             &0<\Delta<1\\
-[1+(-1)^n](2\pi^2)^{-1}\,n^{-2}                  & \Delta=0\\
(-1)^n A\, n^{-4\pi^2 \delta}-\delta n^{-2}   &-1<\Delta<0\\
(-1)^n B\,n^{-1}\sqrt{\ln n} -\delta n^{-2}     &\Delta=-1
\end{cases}
\end{equation}
where we used the shorthand notation
\begin{equation}
\label{delta:def}
\delta=\frac{1}{4\pi \arccos(\Delta)}
\end{equation}

We now insert \eqref{rn} into equation \eqref{ot}. The amplitude $A=A(\Delta)$ has been guessed relatively recently (see \cite{lukyanov99}), yet we do not need this result. Indeed, the leading oscillating term in $\rho^z(n)$ has the exponent $a=4\pi^2 \delta$ varying in the range $1<a<2$ when the anisotropy parameter varies in the $-1<\Delta<0$ range.  Because this leading term oscillates, we form pairs and find that the oscillating terms in (\ref{ot}) yield the contribution
\begin{equation}
\label{An}
A\, n^{-a+1}-A\, (n+1)^{-a+1}\simeq aA\,n^{-a}
\end{equation}
that decays faster than $n^{-1}$. Hence the oscillating term provides merely a constant contribution to the variance $D(t)$ while the sub-leading $n^{-2}$ term results in the leading logarithmically diverging contribution. Therefore
\begin{equation}
\label{Dt:XXZ}
D(t) = \delta \ln t
\end{equation}
This prediction implies that the logarithmic behavior of the variance is universal. 
The amplitude diverges at $\Delta=1$; analytically $\delta\to (4\pi)^{-1}[2(1-\Delta)]^{-1/2}$ as $\Delta\uparrow 1$. This divergence is not very surprising since the isotropic Heisenberg ferromagnet apparently exhibits a truly different behavior. 

In the other extreme $\Delta\downarrow -1$, the amplitude $\delta$ approaches the finite value $(4\pi^2)^{-1}$. However, as indicated by the last formula in \eqref{rn}, the oscillating asymptotic is $Bn^{-1}\sqrt{\ln n}$ for the isotropic anti-ferromagnet \cite{affleck98}. A calculation similar to \eqref{An} gives 
$n^{-2}\sqrt{\ln n}$ after canceling the oscillations. This leads to the following quite surprising behavior:
\begin{equation}
\label{D:AFM}
D\sim \int \frac{dn}{n}\,\sqrt{\ln n} \sim (\ln t)^{3/2}
\end{equation}
Note that for the isotropic anti-ferromagnet the magnetization is non-trivial in the interval that grows slower than linearly with time. (The natural guess is the diffusive $\sqrt{t}$ growth.) However, the upper limit in the integral in \eqref{D:AFM} would affect only the pre-factor. 

Thus our tentative predictions for the variance of the time integrated current are: (i) The enhanced logarithmic growth  $D\sim (\ln t)^{3/2}$ for $\Delta=-1$; (ii) The universal logarithmic behavior $D(t) = \delta \ln t$ with known pre-factor \eqref{delta:def} in the critical region $-1<\Delta<1$.

\section{Discussion}
\label{discussion}

We have studied the fluctuations of the time integrated magnetization current in the quantum $XX$ chain that evolves starting from an inhomogeneous non-stationary initial state \eqref{initial}. We have derived an exact formula \eqref{Dt-exact} for the variance $D(t)$ of the current. We have shown that the variance increases logarithmically in the long time limit \eqref{Dt-asymp}, which is consistent  \cite{disagree} with numerical evaluation of the exact formula (see Fig.~\ref{dtime}). In addition to this logarithmic growth, we have observed oscillations with decreasing amplitude. We have argued that this logarithmic leading order behavior remains unchanged for a more general class of initial conditions (where the magnetization on the two half-chains is not saturated). 

These small logarithmic fluctuations reflect the ideal conductor nature of the integrable $XX$ quantum chain. The current simply ``slides'' through the system ballistically with no disturbance, hence the tiny fluctuations. Conversely, if an impurity is present at the origin, the variance grows linearly with time \cite{schon07}. Similarly, in stochastic particle systems \cite{Ferrari1994, Derrida1998, Lebowitz1999, Johansson2000, Prahofer2000, bodineau04} the noise -- which is intrinsically present in these models -- generates algebraic fluctuations. 

We have argued that current fluctuations in the inhomogeneous $XXZ$ model are also logarithmic (in the critical region). 
Our arguments are heuristic and a more rigorous derivation is a key challenge for future work. Intriguingly, fluctuations seem more tractable than e.g. the average magnetization profile in the $XXZ$ chain, which is completely unknown.

For free fermion systems, current fluctuations were found to be asymptotically Gaussian
\cite{schon07} and therefore the variance provides a complete characterization of the full current statistics.  Moreover, according to \cite{PreKlich2008} this indicates that the entanglement (between the left and right halves) is simply proportional to the variance of the current (with a factor $3/\pi^2$). In order to check this relation we compared our results for $D(t)$ to numerical results for the time dependent entanglement entropy (for the same model and initial condition) presented in Fig.~19 of \cite{gobert05}, and we found a good agreement in the leading order.

Little is known about higher moments of current fluctuations for interacting fermions. There is no reason to believe that they are Gaussian and the full current statistics is very difficult to probe (both theoretically and experimentally) for interesting interacting fermion systems \cite{BB}. (Even for {\em classical} interacting particles, the derivation of the full counting statistics is usually a formidable challenge, see e.g. \cite{Derrida1998, Johansson2000, bodineau04, Sasamoto2005a, Ferrari2006, Rakos2006a, PK}.)
For the $XXZ$ model, however, it could be possible to compute higher order cumulants by employing the heuristic approach which we have applied to computing the variance.

\section{Acknowledgments}

We thank Alexander Abanov, Deepak Dhar, Viktor Eisler, R\'obert Juh\'asz, Eduardo Novais, Pierre Pujol, Zolt\'an R\'acz and Gunter M.\ Sch\"utz for illuminating discussions. We also gratefully acknowledge financial support from 
NIH grant R01GM078986 (T.~A.), 
the Hungarian Scientific Research Fund OTKA PD-72607 (R.~A.), and Jeffrey Epstein for support of the Program for Evolutionary Dynamics at Harvard University.

\appendix

\section{Hydrodynamic description: Examples}
\label{examples}

\subsection{$XX$ model}
\label{examples_XX}

In section \ref{hidro} we demonstrate how the hydrodynamic approach works for the $XX$ model in the simplest case, i.e., when all the spins point up (down) on the left (right) side at $t=0$. One can easily recover the exact results for a slightly more complicated initial state as well. Consider a product state composed of a ground state (in an appropriate external field) with $m_L$ ($m_R$) magnetization on the left (right) side of the chain. That is, the modes $-k_{L(R)}<k<k_{L(R)}$ with $k_{L(R)}=\pi(1/2+m_{L(R)})$ are filled on the left (right) side. Without any restriction, in the following we assume that $m_L>m_R$. 

There are two qualitatively different cases. (1) If the sign of $m_L$ and $m_R$ are equal the scaling profile consists of three segments. Here we assume that $m_L>0, m_R>0$ but the case $m_L<0, m_R<0$ is entirely analogous. Applying the argument of section \ref{hidro} here the magnetization profile takes the following scaling form:
\begin{equation}
m(x) = \left\{
\begin{array}{ll}
 \frac{m_L+m_R}{2} & 0<x<v_L \\
 \frac{m_R}{2}+\frac{1}{4}-\frac{\arcsin(x)}{2\pi}, & v_L<x<v_R \\
 m_R & v_R<x,
\end{array}
\right.
\end{equation}
with $v_L\equiv v(k_L)=\cos(\pi m_L)$ and $v_R\equiv v(k_R)=\cos(\pi m_R)$. 
and $m(-x)=m_L+m_R-m(x)$.

\begin{figure}
\includegraphics{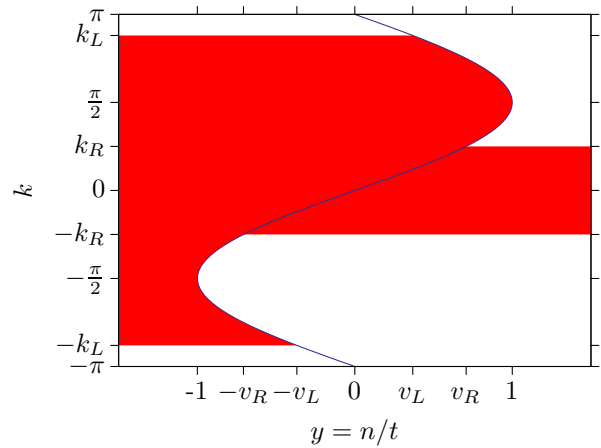}
\caption{(Color online) Hydrodynamic description for the $XX$ chain started from an initial condition where the left (right) side of the chain is in the ground-state with magnetization $m_L$ ($m_R$). We consider the scaling limit where $t\to\infty$ and $y=n/t=\text{const}$. The shaded region shows the elementary excitations that are present in the scaling points indexed by $y$.}
\label{fig:XX_hydro2}
\end{figure}

(2) A qualitatively different case is when $m_L>0$ and $m_R<0$. The argument of section \ref{hidro} leads to the following asymptotic profile:
\begin{equation}
 m(x) = \left\{
\begin{array}{ll}
 \frac{m_L+m_R}{2} & 0<x<v_L \\
 \frac{m_R}{2}+\frac{1}{4}-\frac{\arcsin(x)}{2\pi}, & v_L<x<v_R \\
 m_R+\frac{1}{2} -\frac{\arcsin(x)}{\pi} & v_R<x<1 \\
 m_R & v_R<x,
\end{array}
\right.
\end{equation}
As an illustration see figure \ref{fig:XX_hydro2}. Here we assumed $v_L<v_R$, as the opposite case is entirely similar. In all cases $m(x)$ has the symmetry property $m(-x)=m_L+m_R-m(x)$. We mention that the special case of $m_R=-m_L$ was considered in detail in \cite{spin99,ogata02b}.

\subsection{Dimerized $XX$ model}

\begin{figure}
\includegraphics{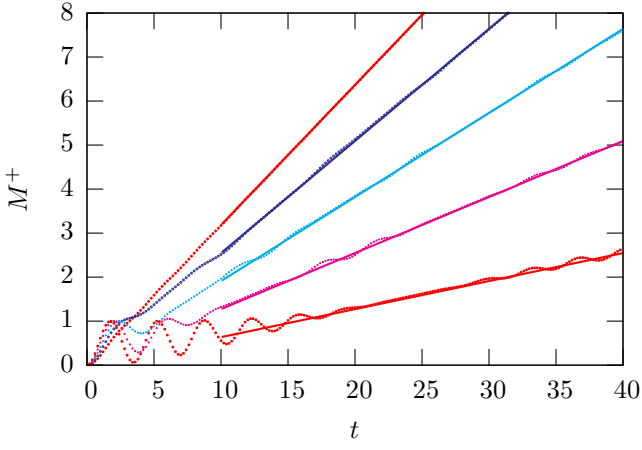}
 \caption{(Color online) Shown is the average transferred total magnetization $M^+=\langle \sum_{n\geq 0} (S^z_n+1/2)\rangle$ as a function of time started from the $|\cdots\uparrow\uparrow\uparrow\downarrow\downarrow\downarrow\cdots\rangle$ initial condition in the dimerized $XX$ model. 
The data points show the result of a time-dependent DMRG calculation published in \cite{gobert05}, solid lines indicate the prediction of the hydrodynamic description.
The coupling constants for the weak link are 1, 0.8, 0.6, 0.4, 0.2 (from above to below).}
 \label{fig:comparison}
\end{figure}

In all the above example the speed of the front was one, but it is not necessary. Consider, as an example, the dimerized $XX$ model \cite{berim02}
\begin{equation}
H = -\sum J_m (s_m^x s_{m+1}^x + s_m^y s_{m+1}^y) 
\end{equation}
with $J_m=1$ for odd $m$, and $J_m=\delta<1$ for even $m$. The spectrum of the model is
\begin{equation}
\label{dimerized_spectrum}
\omega(\tilde k) = \pm \frac{1}{2} \sqrt{1+\delta^2 + 2\delta \cos \tilde k},
\end{equation}
where $\tilde k$ is a sublattice wave number going from $-\pi$ to $\pi$ (an excitation with sublattice wave number $\tilde k$ has momentum $k=\tilde k/2 $). The two signs in (\ref{dimerized_spectrum}) indicate two branches with $k\in(-\pi/2,\pi/2)$. 
For the simplest initial condition with $m_L=-m_R=1/2$,
the fastest mode present on the left [at $\tilde k= \pm(\pi-\arccos \delta)$] has speed $\delta$. That is, the front still moves ballisticly, but it can go arbitrarily slowly. By inverting the function $\omega'(\tilde k)=x/2$
\begin{equation}
k_{1,2} = \arccos \frac{-x^2\pm\sqrt{(1-x^2)(\delta^2-x^2)}}{\delta}
\end{equation}
with $0<x<\delta$, after some elementary calculations, one obtains the magnetization profile as well
\begin{equation}
\label{dimerprof}
  m(x) = -\frac{1}{2} + \frac{1}{2\pi}(k_2-k_1)= -\frac{1}{\pi} \arcsin \left( \frac{x}{\delta} \right ) .
\end{equation}
Surprisingly, the dimerization only rescaled the asymptotic magnetization profile, and the result agrees with a homogeneous chain with coupling $\delta$ at each bond. That means that the weaker links behave as bottlenecks and they govern the dynamics of the chain. 
The average time-integrated current through the origin is $t\delta/\pi$, which was also found to be linear in time in \cite{gobert05}. The value of their current agrees with our result (see figure \ref{fig:comparison}).

\section{Large time asymptotic of $D(t)$}
\label{long}

Here we study the large time behavior of (\ref{Dt-exact}).  
As the summation in \eqref{Dt-exact} goes to infinity, we use a general large $t$ asymptotic formula \cite{gradshteyn} of Bessel functions which is valid for all $m<t$:
\begin{equation}
\label{Bes-asym}
J_m(t) = \sqrt{\frac{2}{\pi\,t}}\,\frac{\displaystyle\cos\left[tf(\mu)-\frac{\pi}{4}\right] }{(1-\mu^2)^{1/4}}
 + \mathcal{O}\left(\frac{1}{t}\right),
\end{equation}
where 
\begin{equation}
\label{f-def}
f(\mu)=\sqrt{1-\mu^2}-\mu\,\cos^{-1}\mu\,,\quad \mu=\frac{m}{t}.
\end{equation}
We will also use the scaled variable $\lambda=l/t$. For $m>t$ the Bessel function $J_m(t)$ becomes exponentially small thus we can safely neglect those terms in (\ref{Dt-exact}). Hence in the double sum we consider only the range $\lambda,\mu\leq 1$. 

Based on (\ref{Bes-asym}) the asymptotic form of a Bessel function with a shifted index can be obtained by simply Taylor expanding $f(\mu)$ in the argument of the cosine (the term coming from the expansion of the denominator is negligible), which gives 
\begin{equation}
\label{Bes-asym-}
 J_{m-1}(t) = \sqrt{\frac{2}{\pi t}}\,\frac{\displaystyle \mu\cos a - \sqrt{1-\mu^2 \sin a} }{(1-\mu^2)^{1/4}}
 + \mathcal{O}\left(\frac{1}{t}\right)
\end{equation}
where we introduce the shorthand notation
\begin{equation*}
a=tf(\lambda)-\frac{\pi}{4}\,,\quad b=tf(\mu)-\frac{\pi}{4}.
\end{equation*}

Now using the asymptotic formulas \eqref{Bes-asym} and \eqref{Bes-asym-} in \eqref{Dt-exact} we obtain
\begin{equation}
\label{bra}
J_{l-1}(t) J_{m-1}(t) + J_l(t) J_m(t)=
\frac{(2/\pi t)T}{(1-\lambda^2)^{1/4}(1-\mu^2)^{1/4}},
\end{equation}
where
\begin{eqnarray}
\label{T}
T &=& (1+\lambda\mu)\cos a\,\cos b
+ \sqrt{1-\lambda^2}\,\sqrt{1-\mu^2}\,\sin a\,\sin b\nonumber\\
&-&\lambda\,\sqrt{1-\mu^2}\cos a\,\sin b-
\mu\sqrt{1-\lambda^2}\,\sin a\,\cos b.
\end{eqnarray}
From hereafter we omit noting the relative $\mathcal{O}(1/\sqrt t)$ corrections. 

When $\lambda,\mu\ll 1$, we can replace $(1-\lambda^2)^{1/4}(1-\mu^2)^{1/4}$ by
one in \eqref{bra}, and the expression for $T$ simplifies to
\begin{equation*}
T=\cos(a-b)=\cos[tf(\lambda)-tf(\mu)].
\end{equation*}
Using (\ref{f-def}) we expand $f(\lambda)$ and $f(\mu)$ to give
\begin{equation*}
tf(\lambda)-tf(\nu-\lambda)=-\frac{\pi(l-m)}{2}+\frac{l^2-m^2}{2t}+ t \mathcal{O}(\lambda^4, \mu^4).
\end{equation*}
When $l, m \ll \sqrt{t}$, we can keep only the first leading term of this sum. In this limit \eqref{bra} reads
\begin{equation}
\label{smalln}
J_{l-1}(t) J_{m-1}(t) + J_l(t) J_m(t)=\frac{2}{\pi t}
\cos\left[\frac{\pi(l-m)}{2}\right].
\end{equation}

Now in \eqref{Dt-exact} we write $\sum_{l,m\geq 1}=\sum_{n\geq 2}\sum_{l+m=n}$, and divide the summation through $n$ into four parts
\begin{equation}
\label{Dterms}
 \sum_{n\geq 2} = \sum_{n=2}^{[t^{1/2-\epsilon}]} + \sum_{n=[t^{1/2-\epsilon}]+1}^{[t^{1/2+\epsilon}]} + \sum_{n=[t^{1/2+\epsilon}]+1}^{[t]} + \sum_{n={[t]} +1}^\infty
\end{equation}
where $[\cdot]$ denotes the integer part. The corresponding terms in \eqref{Dt-exact} will be referred to as 
\begin{equation}
 D(t)=D_1(t)+D_2(t)+D_3(t)+D_4(t)
\end{equation}
respectively. We choose the exponent in \eqref{Dterms} in the range $0<\epsilon<1/2$; later we shall take the $\epsilon\to0$ limit. Since $t^{1/2-\epsilon}\ll t^{1/2}$ as $t\to\infty$, in $D_1(t)$ we can use the simple formula \eqref{smalln}, and immediately perform
one summation. Thus the first sum becomes
\begin{equation}
D_1(t)=\pi^{-2}\sum_{n~ {\rm even}}^{[t^{1/2-\epsilon}]} \frac{1}{n-1},
\end{equation}
which in the $t\to\infty$ limit is simply
\begin{equation}
\label{D1}
D_1(t) = \frac{(1/2-\epsilon)\ln(t)+\gamma_E+\ln(2)}{2\pi^2},
\end{equation}
where $\gamma_E=0.5772\dots$ is the Euler constant.

Now we consider the contribution from the region $t^{1/2-\epsilon}\leq n\leq t^{1/2-\epsilon}$. Using $\cos^2(\cdot)\leq 1$ we find that
\begin{equation}
\label{D2}
D_2(t) \le \sum_{n=[t^{1/2-\epsilon}]+1}^{[t^{1/2+\epsilon}]} \frac{1}{n-1} \approx \frac{2\epsilon}{\pi^2} \ln(t)
\end{equation}
for fixed positive $\epsilon$ and large $t$. Hence one can see that for small $\epsilon$ the contribution from $D_2(t)$ becomes negligible (as compared to $D_1(t)$.

In $D_3(t)$ and $D_4(t)$  we can replace summation by integration, since $t^{1/2+\epsilon}\to\infty$ as $t\to\infty$. We take the square of \eqref{T} and drop rapidly oscillating terms (like $\sin a \cos a$) while $\sin^2$ and $\cos^2$ are replaced by 1/2; this results in replacement of $T^2$ by $(1+\lambda\mu)/2$. Overall, we find that
\begin{multline}
\label{int1}
D_3(t) = \frac{1}{2\pi^2}\sum_{n=[t^{1/2+\epsilon}]+1}^{[t]} \frac{1}{(n-1)^2} \\
\times \sum_{l+m=n} \frac{1+\lambda\mu}{\sqrt{(1-\lambda^2)(1-\mu^2)}}
\end{multline}
and
\begin{multline}
\label{int2}
D_4(t) = \frac{1}{2\pi^2}\sum_{n=[t]+1}^{[2t]} \frac{1}{(n-1)^2} \\
\times \sum_{l+m=n} \frac{1+\lambda\mu}{\sqrt{(1-\lambda^2)(1-\mu^2)}}.
\end{multline}
Note that there is only an exponentially small contribution from $n>2t$ terms.

Notice that the second sum in (\ref{int1}-\ref{int2}) can be replaced by
\begin{equation}
 \frac{1}{n}\int_0^1 F(\nu,x) dx,
\end{equation}
where we used the shorthand notation
\begin{equation}
\label{Fnx}
F(\nu,x) = \frac{1+\nu^2 x(1-x)}{\sqrt{[1-\nu^2 x^2][1-\nu^2(1-x)^2]}},
\end{equation}
and introduced $\nu=n/t=\lambda+\mu$ and $x=\lambda/\nu$.
For $t\to\infty$ the first sum in (\ref{int1}-\ref{int2}) can also be replaced by an integral, which leads to
\begin{equation}
\label{integral}
D_3(t) = \frac{1}{2\pi^2}\int_{t^{\epsilon-1/2}}^1 \frac{d\nu}{\nu}
\int_0^1 dx\,F(\nu,x)
\end{equation}
It is useful to rewrite (\ref{integral}) as 
\begin{multline}
D_3(t)=\frac{1}{2\pi^2}\int_{t^{\epsilon-1/2}}^1 \frac{d\nu}{\nu}\int_0^1 dx\,[F(\nu,x)-1] \\ + \frac{(1/2-\epsilon)\ln t}{2\pi^2}, 
\end{multline}
where the first integral is convergent in the limit $t\to\infty, \epsilon\to 0$. Therefore the contribution gathered in the $t^{1/2+\epsilon}\leq n\leq t$ region is 
\begin{equation}
\label{D3}
D_3(t) =  \frac{(1/2-\epsilon)\ln t + C_3}{2\pi^2}
\end{equation}
where
\begin{equation}
C_3=\int_0^1 \frac{d\nu}{\nu}
\int_0^1\! dx\,[F(\nu,x)-1] = 0.34929294\ldots
\end{equation}

By a similar argument the contribution \eqref{int2} from the $t\leq n\leq 2t$ region remains finite in the $t\to\infty$ limit:
\begin{equation}
\label{D4}
 D_4(t) = \frac{C_4}{2\pi^2}
\end{equation}
with
\begin{equation}
C_4 = \int_1^2 \frac{d\nu}{\nu}
\int_{1-\nu^{-1}}^{\nu^{-1}} dx\,F(\nu,x) = 1.34385423\ldots
\end{equation}
Combining the contributions \eqref{D1}, \eqref{D2}, \eqref{D3}, and \eqref{D4} from the three regions of \eqref{Dterms}, and taking now the $\epsilon\to0$ limit, we obtain
\begin{equation}
D(t)=\frac{1}{2\pi^2}(\ln t + C)
\end{equation}
where
\begin{equation}
\label{C}
C = \gamma_E + \ln 2 + C_3+ C_4 = 2.963510026\ldots
\end{equation}



\begin{thebibliography}{99}
\bibitem{disagree}
    The long time asymptotics of current fluctuations for the free fermion case 
    were also considered in Ref.~\cite{schon07}. For the second cumulant, 
    a logarithmic growth was found with prefactor twice larger than in 
    Eq.~\eqref{Dt-asymp}.
\end{thebibliography}
\end{document}